\documentclass[figures]{epl}
\title{Spin-Polarization transition in the two dimensional electron gas}
\shorttitle{Spin-Polarization transition in etc.}
\author{D. Varsano\inst{1} \and  S. Moroni\inst{1}\and  G. Senatore\inst{2} }
\institute{
          \inst{1} INFM, Universit\`a di Roma ``La Sapienza'' - 
                       P.le Aldo Moro 2, 00185, Roma, Italy\\
          \inst{2} INFM, Dipartimento di Fisica Teorica, Universit\`a di Trieste - 
                       Strada Costiera 11, 34014, Trieste, Italy
         }

\begin{document}

\maketitle

\begin{abstract}
We present a numerical study of magnetic phases of the 2D electron gas near
freezing. The calculations are performed by diffusion Monte Carlo in the fixed
node approximation. At variance with the 3D case we find no evidence for the
stability of a partially polarized phase. With plane wave nodes in the trial
function, the polarization transition takes place at $r_s=20$, whereas the
best available estimates locate Wigner crystallization around $r_s=35$.  Using
an improved nodal structure, featuring optimized backflow correlations, we
confirm the existence of a stability range for the polarized phase, although
somewhat shrunk, at densities achievable nowadays in 2 dimensional hole
gases in semiconductor heterostructures . The spin susceptibility of the
unpolarized phase at the magnetic transition is approximately 30 times the
Pauli susceptibility.
\end{abstract}

Quantum simulations show that the three--dimensional electron gas undergoes a
continuous spin-polarization transition, upon decreasing the density into the
strongly correlated regime, before forming a Wigner crystal~\cite{cep82,ohb}.
Despite the theoretical interest of a simple model exhibiting quantum phase
transitions~\cite{girvin}, the difficulty of realizing a low density electron
gas in real materials makes the contact with experiment a rather indirect
one~\cite{young}.
However, electrons can be confined into effectively two--dimensional systems,
for instance in Si MOSFET's and III-V semiconductor
heterostructures~\cite{ando}, over a density range extending down to the
freezing transition~\cite{yoon}. 

Far from being a mere playground for testing many--body theories and numerical
simulations, strongly correlated two--dimensional electronic systems offer an
extremely rich and interesting phenomenology, such as the fractional quantum
Hall effect~\cite{tsui} and a previously unexpected metal--insulator
transition~\cite{kravchenko}, not to mention their relevance to superconductor
cuprates~\cite{bednorz}.  Therefore, we think that it is useful to extend the
work of ref.~\cite{ohb} and study the polarization transition of the
two--dimensional electron gas (2DEG). In fact, spin fluctuations appears to
play an important role in the 2DEG at the metal--insulator transition in
presence of disorder~\cite{belitz}).

Previous quantum Monte Carlo (QMC)
simulations~\cite{tanatar,mcs,kwon,rapisarda,cornell} of the 2DEG provide
little information on the magnetic properties of the system, with only one
spin susceptibility calculation\cite{cornell} 
at non--zero wave vectors and relatively high densities
($r_s\leq 10$, where $r_sa_B=1/\sqrt{\pi\rho}$ with $a_B$ the Bohr radius and
$\rho$ the density), and one attempt at exploring the partially polarized
case~\cite{tanatar} near freezing.
We emphasize that we have only mentioned, here,  results obtained from
QMC, a computationally demanding numerical method which 
projects the ground state from a trial function $\Psi_T$ by a stochastic
technique~\cite{qmc_reviews}. Both in this work and in 
refs.~\cite{ohb,tanatar,mcs,kwon,rapisarda,cornell} 
the method is implemented in the fixed--node (FN)
approximation~\cite{fna}, giving the lowest upper bound to the ground
state energy consistent with the {\it nodal structure} of $\Psi_T$.
Though not exact unless $\Psi_T$ has the same nodes
as the unknown ground state, a condition not generally met, FN--QMC 
provides fairly accurate predictions for ground state 
properties of interacting fermion systems.
Other approximate theories may fulfill such an accuracy requirement
at weak coupling~\cite{giuliani}, but are not amenable to 
systematic improvement in the strongly correlated regime.
When applied beyond their range of validity, they
predict a variety of magnetic instabilities at 
unrealistically high densities~\cite{sato_and_pathak}.

\begin{figure}[t]
\setlength{\abovecaptionskip}{-0.5cm}
\vspace{-0.5cm}
\onefigure[scale=0.4]{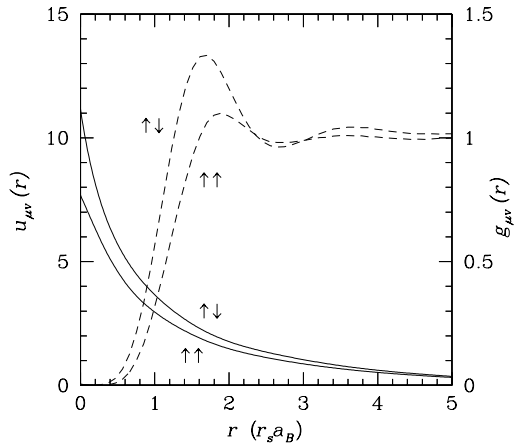}
\caption[]{Optimized pseudopotentials $u_{\mu\nu}(r)$ (solid lines,left scale) 
and VMC pair distribution functions $g_{\mu\nu}(r)$ (dashed line, right scale)
for the unpolarized 2DEG at $r_s=20$.}
\label{fig:pseudo}\end{figure}

We simulate 2D systems of $N=N_\uparrow+N_\downarrow$ electrons, interacting
with a pair potential $v(r)=2/(r_sr)$, in a square box of side $L=\sqrt{N\pi}$
with periodic boundary conditions (energies are in Ry and distances in
$r_sa_B$ units).  A rigid uniform background of positive charge ensures
neutrality.  The polarization is $\zeta=(N_\uparrow-N_\downarrow)/N$.
Following ref.~\cite{ohb}, we use a Jastrow--Slater form for the trial
function, $\Psi_T=\exp[-\sum_{i<j}u_{\mu\nu}(r_{ij})]D_\uparrow D_\downarrow$.
Here the indices $i,j$ label electrons, $\mu$ and $\nu$ label the spin of
particles $i$ and $j$; $u(r)_{\mu\nu}$ are pair pseudopotentials for like
($\mu=\nu$) or unlike--spin ($\mu\ne\nu$) particles, optimized for each system
considered, and $D_{\uparrow(\downarrow)}$ is a Slater determinant of plane
waves for up(down)--spin electrons. 
As it has been previously done in three dimension \cite{cep82,ohb}, 
we are only considering homogeneous phases\cite{sdw}.

Figure~\ref{fig:pseudo} shows
the optimized pseudopotentials for an unpolarized system with $N=114$
at $r_s=20$. In the same figure, the spin--resolved pair distribution
functions $g_{\uparrow\uparrow}(r)$ and $g_{\uparrow\downarrow}(r)$,
computed with variational Monte Carlo (i.e. by taking the expectation value
over $|\Psi_T\rangle$) is shown by the dashed lines.
Despite the fact that the like--spin pseudopotential is less repulsive,
the pair distribution functions do not show the local ferromagnetic order
predicted~\cite{ohb} in 3D across the magnetic transition. Note that in 2D, at 
the accuracy level of FN--QMC with plane wave determinants, the energies 
of the unpolarized and the fully polarized phases cross precisely at 
$r_s=20$, according to both ref.~\cite{rapisarda} and the present work
(see below).

\begin{figure}[t]
\setlength{\abovecaptionskip}{-0.5cm}
\vspace{-0.5cm}\hspace{-0.7cm}
\twoimages[scale=0.4]{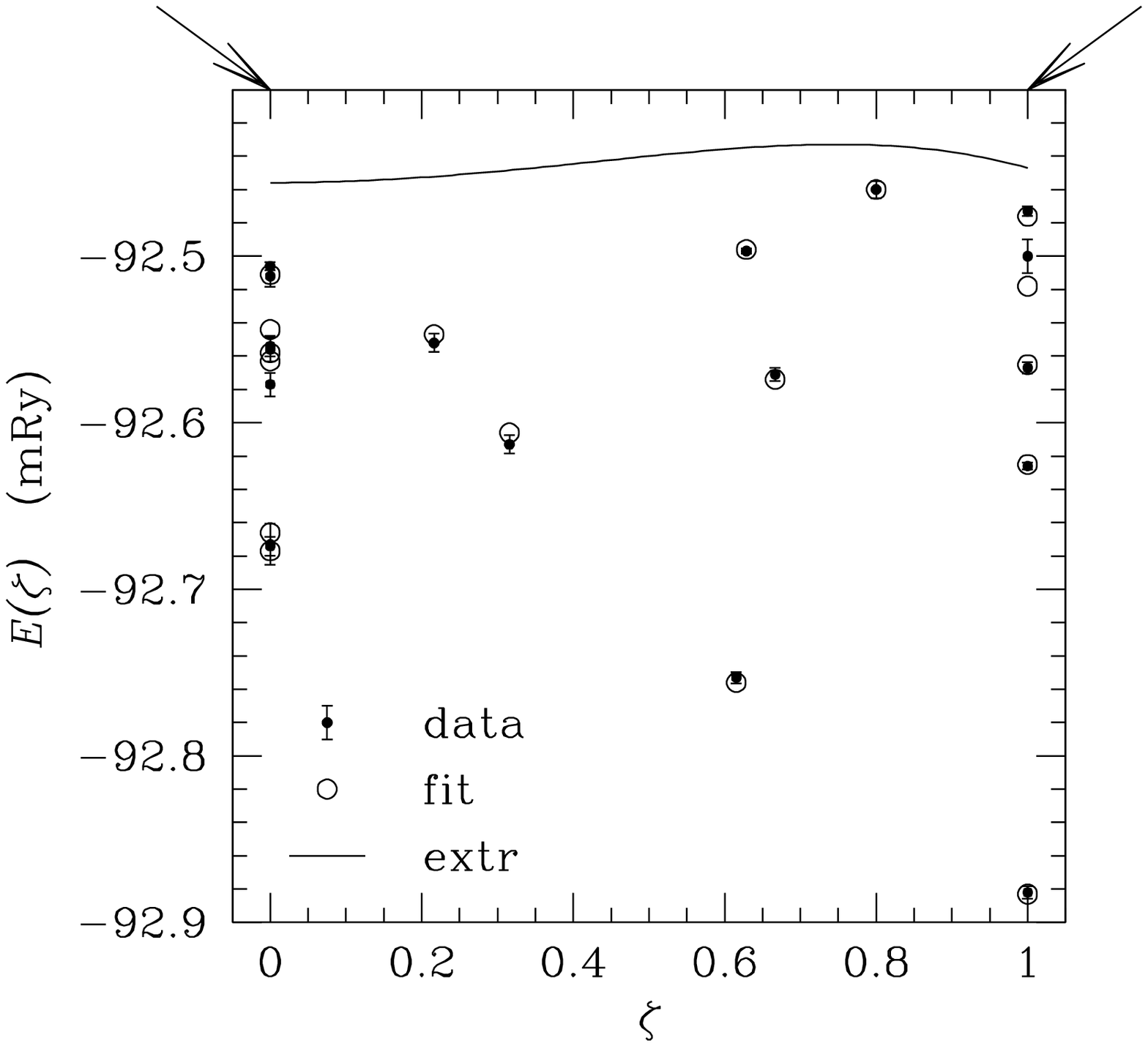}{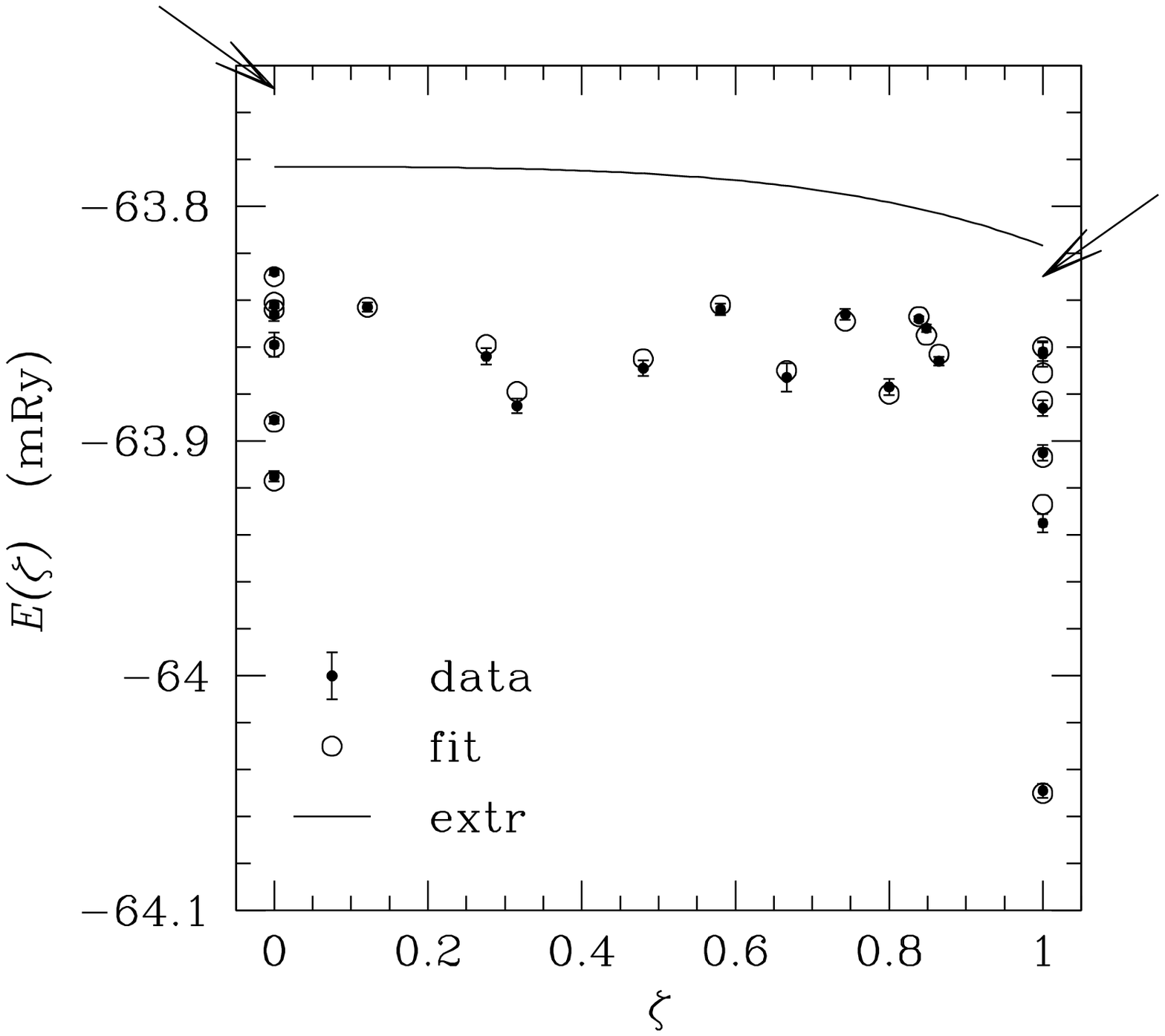}
\caption[]{Plane wave fixed node energy versus polarization at $r_s=20$ (left
panel) and at $r_s=30$ (right panel).}
\label{fig:rs20-30}
\end{figure}

Our estimate of energy vs. polarization in the thermodynamic limit,
parametrized in the form
$E(\zeta)=E(\zeta=0)+\beta\zeta^2+\gamma\zeta^4$, is shown by the
solid line for $r_s=20,30$, in fig.~\ref{fig:rs20-30}.
The statistical uncertainty is less than $10^{-5}$.
The points with errorbars are FN--QMC results with plane waves, and the open 
circles are the result of the fit for the number extrapolation (see below).
The arrows indicate the results of ref.~\cite{rapisarda} at $\zeta=0$ 
and $\zeta=1$.

The polarized and unpolarized phases have nearly the same energy at $r_s=20$,
whereas at $r_s=30$ the polarized fluid is stable. We find no evidence for the
stability of a partially polarized phase at these densities, and we believe
that this is the case for other densities as well (a reentrant partially
polarized phase could form for $r_s < 20$; alternatively, $\gamma$ should
change sign twice\cite{note1} between $r_s=20$ and $r_s=30$: both cases seem
unlikely).  This conclusion agrees with a conjecture of ref.~\cite{tanatar},
based on the simulation of a single partially polarized system at $r_s=40$;
however, 
the more systematic analysis presented here
seems worth the effort, also in view of some discrepancies existing 
between the results of ref.~\cite{tanatar} and later 
simulations~\cite{kwon,rapisarda}.

The dispersion of the FN $E(\zeta)$ is very weak, which reflects in
a large spin susceptibility $\chi_s$ of the unpolarized 2DEG at $r_s=20$, 
where the first--order polarization transition takes place. The best--fit 
value of $\beta$ is 8.2$\pm 1.2\times$10$^{-5}$, corresponding to 
$\chi_s/\chi_P\simeq 30$, where $\chi_P$ is the Pauli susceptibility. 
This is an extremely large value of $\chi_s$. We note that a 
phenomenological model for $\chi_s$~\cite{iwamoto}, 
based on including known sum rules and fitting QMC
data from ref.~\cite{tanatar}, gives $\chi_s/\chi_P=1.6$ at $r_s=20$.

\begin{table}[t]
\caption[]{Energy--fitting parameters from DMC. See text for the meaning of
symbols}
\label{tab:dmcfit}
\begin{center}
\begin{tabular}{c|cccccccc}
& $E(\zeta=0)$ & $\beta\times 10^5$ & $\gamma\times 10^5$ & 
       $\Delta$ & $\epsilon$ & $b$ & $\alpha$ &$\chi^2$ \\
\hline
$r_s=20$&-0.092456&8.22&-7.34&0.00129&2.0&0.00712&-0.0030&1.7\\
$r_s=30$&-0.063783&-0.64&-2.72&0.00045&1.8&0.00397&-0.0007&1.9\\
\end{tabular}
\end{center}
\end{table}

We now discuss the differences between our results
and the ones of refs.~\cite{rapisarda,kwon} (all three calculations 
are nominally equivalent).  We perform QMC simulations in the Diffusion 
Monte Carlo (DMC) version~\cite{dmc}.  A population of $N_W$ walkers 
(copies of the system) perform a random walk in configuration space. 
The time steps of the random walk stochastically follow 
a short time approximation to the imaginary time evolution of the quantum 
system.  The results need to be extrapolated to zero time step $\tau$
and, in the implementation used, to infinite $N_W$, as well as to
the standard thermodynamic limit $N\to\infty$.  In brief, (i) we 
perform the number extrapolation directly on several DMC energies 
instead of sticking the VMC size corrections on the DMC energy computed 
for a single system size~\cite{rapisarda,kwon}, and (ii) we believe that 
we better control the finite $\tau$ and finite $N_W$ biases.

We fit the energies computed at various polarizations $\zeta$ and particle
numbers $N$ using
\begin{equation}
E_N(\zeta)=E(\zeta=0)+\beta\zeta^2+\gamma\zeta^4
+\Delta(1+\epsilon\zeta^2)(T_N^{(0)}(\zeta)-T^{(0)}(\zeta))
-(b+\alpha\zeta^2)/N,
\end{equation}
where $T_N^{(0)}$ and $T^{(0)}$ are the kinetic energies of a non--interacting
2DEG at $r_s=1$ for $N$ particles and in the thermodynamic limit, respectively.
The fitting parameters from FN energies with plane
wave nodes are listed in table~\ref{tab:dmcfit}.
Obviously, the fit to the VMC energies yields different values for
$E(\zeta=0)$, $\beta$ and $\gamma$.  Less obvious is the difference of the
other fitting parameters (i.e.  is the number extrapolation the same for DMC
as for VMC?).  In table~\ref{tab:vmcfit} we list the fitting parameters
obtained from VMC (103 systems at each density using parameter-free RPA
pseudopotentials~\cite{cep78} to avoid spurious effects from different levels
of optimization at different particle numbers).
This VMC--DMC comparison may be in part biased by the fact that 
we used Ewald sums~\cite{natoli} for the Coulomb potential in VMC, and a 
truncated potential~\cite{fraser} in DMC. However 
we note that $b$, the fitting parameter directly related to the potential
energy, is nearly the same in VMC and DMC, a significant difference 
being instead present in the kinetic energy terms.

\begin{figure}[t]
\setlength{\abovecaptionskip}{-0.5cm}
\vspace{-0.5cm}
\onefigure[scale=0.4]{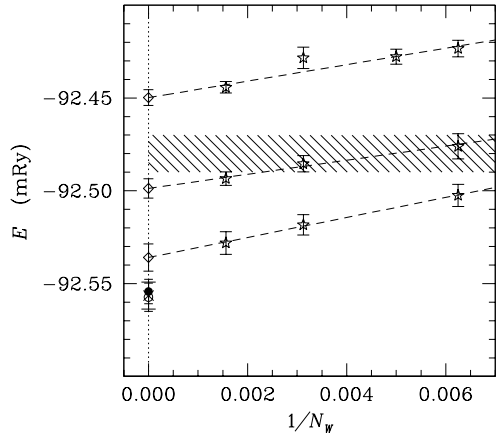}
\caption[]{Effects of finite time step and number of walkers on the 
fixed node energy for 58 particles at $r_s=20$, in the normal fluid.  See text for the meaning of
symbols}
\label{fig:n58}
\end{figure}

\begin{table}[b]
\caption[]{Energy--fitting parameters from VMC. See text for the meaning of
symbols }
\label{tab:vmcfit}
\begin{center}
\begin{tabular}{c|cccccccc}
& $E(\zeta=0)$ & $\beta\times 10^5$ & $\gamma\times 10^5$ &
       $\Delta$ & $\epsilon$ & $b$ & $\alpha$ &$\chi^2$ \\
\hline
$r_s=20$&-0.091553&-30.2&-60.7&0.00258&1.0&0.00687&0.0026&2.3\\
$r_s=30$&-0.063237&-52.3&-43.0&0.00090&2.1&0.00355&0.0009&2.8\\
\end{tabular}
\end{center}
\end{table}

We turn now to the time step error and the population control bias.
The results for $r_s=20$ shown in fig.~\ref{fig:rs20-30} are all 
extrapolated at infinite number of walkers
and zero time step.
For $r_s=30$ the extrapolation has been done in 10 cases; then
a time step ${\tilde \tau}$ and a number of walkers ${\tilde N_W}$
have been chosen such that the residual bias is less than $5\times 10^{-6}$
in the selected cases. Some of the results shown in fig.~\ref{fig:rs20-30} 
for $r_s=30$ are obtained using ${\tilde \tau}$ and ${\tilde N_W}$.

For 58 particles at $r_s=20$, both ref.~\cite{rapisarda} and~\cite{kwon} 
report $E=-0.09248(1)$.
However we find that the extrapolation at
zero time step and infinite number of walkers is $-0.092557(76)$.
In fig.~\ref{fig:n58} the results for various numbers of walkers and 
time steps are shown by star symbols. For each $\tau$ the 
extrapolation (linear in $1/N_W$) to infinite number of walkers
is shown by the diamond. 
The extrapolation (linear in $\tau$) of the diamonds to zero time 
step is shown by the open circle. The shaded region encloses the
energy values between $-0.09248\pm 1$, i.e. it represents the
result of refs.~\cite{rapisarda,kwon}.

The filled circle indicates a result, extrapolated 
to $\tau=0$ and $N_W=\infty$, obtained with
a different DMC code, in which walkers carry a weight for a while 
and branch every so often~\cite{sorella} instead of 
branching at each time step. 
The cross is the result, extrapolated to $\tau=0$, of yet another
calculation with reptation QMC~\cite{snakes}, 
a single--thread algorithm (hence, one having no population control bias).
The finite--$\tau$ and finite--$N_W$ biases are all different in 
these calculations, but the extrapolated result is the same.
This cross--check performed using three different algorithms gives us 
confidence on our extrapolated results.

The above analysis demonstrates the importance of an extremely 
accurate control of the convergence parameters of the numerical 
calculations, due to the relevance of small energy differences 
on physical properties related to spin polarization in this 
strongly coupled system.

We finally consider the FN approximation. In 3D, the error
of the FN energies with plane wave nodes is presumably 
negligible~\cite{ohb} in characterizing the continuous 
polarization transition. On the energy scale of 
the weak dependence of $E$ on $\zeta$ shown in figs.~\ref{fig:rs20-30},
however, the FN bias might be relevant.
We have computed FN energies with better nodes provided
by optimized backflow correlations~\cite{kevin}, at $r_s=20$ 
and $30$ for zero and full polarization. Where available, 
comparison between the FN energy with backflow nodes and 
the exact result~\cite{kwon_te} supports the accuracy of the former for
the 2DEG. Including backflow correlations, 
the FN energy decreases more for $\zeta=0$ than for $\zeta=1$:
the relative gain of the unpolarized phase is found to be
$7.2(7)\times 10^{-5}$ at $r_s=20$ and $2.0(9)\times 10^{-5}$ 
at $r_s=30$. This makes the unpolarized phase stable at $r_s=20$, 
but at $r_s=30$ the energy of the polarized fluid remains lower 
(compare with fig.~\ref{fig:rs20-30} and table~\ref{tab:dmcfit}), even though 
by an amount close to the statistical uncertainty\cite{kwon-unp}. 

At $r_s=30$ the system is still fluid. Based on plane wave FN energies,
crystallization occurs at a lower density, $r_s=37\pm 5$ and $34\pm 4$
according to refs.~\cite{tanatar} and~\cite{rapisarda}, respectively. In fact,
backflow correlations will further stabilize the fluid phases, likely yielding
a  slightly  lower crystallization density (for
the polarized liquid at $r_s=30$ we find a tiny difference of $1.6(4)\times
10^{-5}$ between the plane wave and the backflow FN energy).

Summarizing, we have studied the polarization transition 
in the 2DEG using the FN--DMC method. With plane wave nodes,
the transition is first order and takes place at $r_s=20$,
where the spin susceptibility reaches 30$\chi_P$.
Upon improvement of the nodal structure, obtained including
backflow correlations, the transition density approaches $r_s=30$,
still leaving a stability window for the fully polarized phase between
the paramagnetic fluid and the Wigner crystal.

\acknowledgments
We acknowledge support from INFM under the Parallel Computing Initiative and
from  MURST under the PRIN 1999 Initiative.


\begin{thebibliography}{0}
\bibitem{sdw}
We are not considering the possibility of spatial modulations of either charge
or spin, i.e., we are disregarding here charge and spin density waves.
\bibitem{note1} 
Note that with the chosen form for
$E(\zeta)$, an absolute minimum at $0<\zeta^*<1$ requires: $\beta<0$,
$\gamma>0$, and $\gamma> -\beta/2$
\end{thebibliography}
\end{document}